\documentclass[a4paper,onecolumn]{article}
\usepackage{fullpage}
\usepackage{times}
\usepackage{epsfig}
\usepackage{amsfonts}
\usepackage{amsmath}
\usepackage{amssymb}
\usepackage{amsthm}
\usepackage{color}
\usepackage{multirow}
\usepackage[normalem]{ulem}
\newcommand{\stkout}[1]{\ifmmode\text{\sout{\ensuremath{#1}}}\else\sout{#1}\fi}
\usepackage{latexsym}
\usepackage{mathrsfs}
\usepackage[numbers]{natbib}
\usepackage{verbatim}
\usepackage[utf8]{inputenc}
\usepackage[T1]{fontenc}
\usepackage{float}
\usepackage{dsfont}

\usepackage{mathtools}
\usepackage{textcomp,gensymb}

\usepackage{caption}
\usepackage{subcaption}

\usepackage{graphicx}
\usepackage[table,x11names]{xcolor}

\definecolor{light-gray}{rgb}{0.78,0.78,0.78}

\newcommand\BlackCell[1]{%
  \multicolumn{1}{c|}{\cellcolor{light-gray}\textcolor{black}{#1}}
}

\usepackage{csvsimple}

\usepackage{IEEEtrantools}

\usepackage[colorlinks=true,linkcolor=blue,citecolor=magenta,urlcolor=blue]{hyperref}

\usepackage{enumitem}
\usepackage{varwidth}

\usepackage{bbm} 

\usepackage{cleveref}

\usepackage{authblk}

\title{Quantum-inspired classification\\ based on quantum state discrimination}

\author[1]{Emmanuel Zambrini Cruzeiro}
\author[2]{Christine De Mol}
\author[3]{Serge Massar}
\author[3]{Stefano Pironio}

\affil[1]{Instituto de Telecomunicações,
Lisbon, 1049-001, Portugal}
\affil[2]{Department of Mathematics and ECARES, CP 217, Universit\'e libre de Bruxelles (ULB), Belgium}
\affil[3]{Laboratoire d'Information Quantique (LIQ) CP 224, Universit\'e libre de Bruxelles (ULB), Belgium}

\begin{document}
\maketitle
\begin{abstract}
  We present quantum-inspired algorithms for classification tasks inspired by the problem of quantum state discrimination. By construction, these algorithms can perform multiclass classification, prevent overfitting, and generate probability outputs. While they could be implemented on a quantum computer, we focus here on classical implementations of such algorithms. The training of these classifiers involves Semi-Definite Programming. We also present a relaxation of these classifiers that utilizes Linear Programming (but that can no longer be interpreted as a quantum measurement). Additionally, we consider a classifier based on the Pretty Good Measurement (PGM) and show how to implement it using an analogue of the so-called Kernel Trick, which allows us to study its performance on any number of copies of the input state. We evaluate these classifiers on the MNIST and MNIST-1D datasets and find that the PGM generally outperforms the other quantum-inspired classifiers and performs comparably to standard classifiers.  
\end{abstract}


\section{Introduction}

In recent years, quantum-inspired classical algorithms have emerged as a promising approach for solving various problems \cite{da2006quantum,chia2019quantum}, including those in machine learning \cite{tang2019quantum}. These algorithms draw inspiration from the principles of quantum mechanics but can be implemented on classical computers. In this paper, we explore the potential of using quantum state discrimination as a basis for developing new classical supervised learning methods.

Quantum state discrimination refers to the problem of identifying which one of a known set of quantum states has been prepared based on the outcome of a quantum measurement on the state. Many strategies have been developed to address this problem \cite{Helstrom1969, holevo2011probabilistic, Barnett2009}. On the other hand, supervised machine learning is a subfield of machine learning where the algorithm is trained using labeled examples, and the goal is to learn a mapping between data and labels given the set of training examples. In this paper, we establish a connection between quantum state discrimination and supervised machine learning using the following basic idea: (i) map the training data into quantum states, (ii) find a quantum measurement that best discriminates between the average states corresponding to each of the possible labels, and (iii) predict the label for new input data by carrying out the measurement on the corresponding state. If one implements this on a quantum computer, one has a novel instance of quantum state discrimination. But if one implements this procedure on a classical computer, i.e. by representing the quantum state and the measurement on a classical computer, then one has a novel class of quantum-inspired machine learning algorithms. 

The resulting quantum-inspired classification is interesting for several reasons. First, it is naturally multiclass because quantum measurements can have multiple outcomes, unlike most traditional classification algorithms, which are binary. Second, it outputs not only a unique label but also a probability for each possible label, which can be useful for some applications. Third, it avoids problems related to overfitting because the quantum measurements are described by unit-trace positive operators, which places constraints on the classifier.

In this article, we study such quantum-inspired classification algorithms and present numerical implementations. After building the quantum states associated with the dataset of interest, there are many different quantum measurements that could be carried out on the state and it is not obvious which will be the best to classify the data. We explore various strategies for finding the best quantum measurement, including those based on semidefinite programming and the Pretty Good Measurement (PGM) \cite{belavkin1975optimalA,belavkin1975optimalB,kholevo1979asymptotically,Hausladen1994}. We also consider how the measurement performs on multiple copies of the quantum state. Increasing the number $m$ of copies in principle improves quantum state discrimination and hence could be expected to lead to better classifiers, but the computational cost of finding and implementing the discriminating measurement increases exponentially with $m$. Interestingly, we will show how one can transpose the so-called Kernel Trick \cite{Cortes1995,Hofmann2008} to the PGM, so that the computational cost of implementing the PGM scales with the number of examples in the dataset, rather than with the dimension of the Hilbert space. This allows us to study the performance of the PGM classifier on tensor products of the quantum state, $\rho(u)^{\otimes m}$, with an arbitrary number $m$ of products, and even to study the limit $m \to \infty$. Finally, we provide experimental results for our approach on the MNIST and MNIST-1D datasets and compare its performance to that of classical classifiers based on Ridge Regression and Logistic Regression.

We note that the general connection between quantum state discrimination and supervised classification proposed here has also been introduced in other works, albeit not as extensively as in the present paper. The binary classification case was addressed in \cite{Sergioli2018,Sergioli2019}, while the multiclass case has been explored in \cite{Tiwari2018, giuntini2021quantum, Giuntini2023}. One original contribution of our work not present in the above papers is the application of the Kernel Trick to the PGM, which, as said before, enables us to scale the computational cost with the number of examples in the dataset rather than with the dimension of the Hilbert space.

We also note that a related research direction is the development of quantum algorithms for machine learning which would, contrary to the work presented here, be implemented on a quantum computer. For some representative works in this directions, see for instance the following theoretical \cite{PhysRevA.102.032420,PhysRevA.103.032430,abbas2021power,PhysRevResearch.3.L032049} and  experimental  \cite{grant2018hierarchical,havlivcek2019supervised,tacchino2020quantum} contributions.

This paper is organized as follows. In Section 2, we provide a brief overview of supervised learning, and in Section 3, we introduce the basics of quantum state discrimination. In Section 4, we present our approach to quantum-inspired classification algorithms and describe various strategies for choosing the best measurement. In Section 5, we focus on the PGM and present our application of the Kernel Trick. In Section 6, we present the results of our experiments on the MNIST and MNIST-1D datasets. Finally, we conclude with a brief discussion.

\section{Machine learning and supervised classification}\label{Sec:2}

\paragraph{Input data.}
A supervised machine learning algorithm receives as input $N$ $p$-dimensional feature vectors $u_i\in \mathbb{R}^{p}$, ($i=1,\dots,N$), which we consider as column vectors. To each input $i$ we associate a discrete  label $y_i\in \{1,...,K\} $. 
 The labels thus divide the data into $K$ classes corresponding to the sets $C_k = \{u_i : y_i = k\}$ of data vectors $u_i$ whose associated labels $y_i$ take the  value $k$. We denote $N_k=\vert C_k \vert $ the cardinality of the set $C_k$. We introduce $p_k=N_k/N$, which we can view as the probability of an input having label $k$.

We can represent the input data  as a  $p\times N$ matrix $M$, which we call the dataset, and the labels as a $N\times 1$ column vector $y$.

\paragraph{Task.}
The task of a classification supervised machine learning algorithm is to learn how to correctly predict labels for data that the machine has never seen before. The algorithm learns on a training dataset $M^\text{train}$ and its associated labels $y^\text{train}$. The output of the training procedure is a classifier $f:\mathbb{R}^{p}\to \{1,\dots ,K\}$, i.e. a function that maps data to prediction labels. The performance of the classifier is evaluated on a new dataset $M^\text{test}$, by comparing the correct labels $y^\text{test}$ to the predicted labels $\hat{y}$ obtained by applying $f$ to the dataset $M^\text{test}$.

\paragraph{Preprocessing.}

In the present work we consider classifiers $f$ with the following structure.
First, there may be some preprocessing of the input data through a function
\begin{equation}
X: \mathbb{R}^{p} \to \mathbb{R}^{q} : u_i \to x_i=X(u_i)
\label{Eq:Preprocessing}
\end{equation}
that maps each $p$-dimensional input vector $u_i$ to a $q$-dimensional feature vector $x_i=X(u_i)$.

Examples of preprocessing include \emph{centering}, that is removing the mean of each component of $u_i$,
\begin{equation}
  u_i \to x_i= u_i -E[u]\ ,
\label{Eq:Center}
\end{equation}
where $E[u]=\frac{1}{N}\sum_{i=1}^N u_i$ is the $p$-dimensional vector of the means of the components of $u_i$ computed over the entire dataset. 

 Another simple preprocessing is \emph{standardization}, that is rescaling each component of $u_i$ by their standard deviation,
\begin{equation}\label{Eq:Center2}
u_i \to x_i = \frac{u_i}{\sqrt{ E[u^2] - E[u]^2}}\ .
\end{equation}
This is particularly useful when the different components of $u_i$ have very different standard deviations, or different units. After this operation $u_i$ is dimensionless.

A useful transformation that reduces the number of features is \emph{Principal Component Analysis} (PCA), 
\begin{equation}
  u_i \to x_i= \Pi^{\text{PCA}}_{p'}  u_i
  \end{equation}
  where $\Pi^{\text{PCA}}_{p'}$ projects the original $p$-dimensional vector onto the lower-dimensional subspace corresponding to the $p'<p$ largest principal components.
  
  Finally, one can also increase the number of features, through, e.g. \emph{Polynomial Feature Expansion},  a process in which new features are created by raising the existing features to a power, allowing for non-linear relationships to be captured in a linear model. For instance, the following map
\begin{equation}
u_i \to x_i=(u_i, u_i \otimes u_i)\ ,
\label{Eq:PolyExp}
\end{equation}
builds all monomials of degree 1 and 2 in the components of $u_i$ and increases the number of distinct features from $p$ to $p+p(p+1)/2$. 

\paragraph{Classifiers.}

Given the preprocessed input $X(u_i)$, one then typically computes a score for each class, in a way that has been determined using the training set. 

In the present work we use as score \textit{linear predictor functions} given by 
\begin{equation}
 \text{score}(x_i,k) = \beta_k^T  x_i
\label{Eq:scorebeta_k}
\end{equation}
where $\beta_k$ is the weight vector assigned to class $k$, whose components have been determined by the training procedure.

Finally, to classify a new example/sample to come $z$ (and in particular the samples belonging to the test set), we pick the class with the highest score and assign the label
 \begin{equation}
 \hat y=\text{argmax}_{k}\left( \text{score}(X(z),k)\right) \ .
 \label{Eq:argmaxy}
 \end{equation}

\paragraph{Evaluating a classifier's performance.}
For each pair of classes $k$ and $k'$, one keeps count of the number $V_{k'k}$ of objects $u_i\in C_k$ (or $x_i\in C_k$) belonging to class $k$ which are classified as belonging to class $k'$. The matrix $V$ is called the confusion matrix. From the confusion matrix, several  figures of merit for classification may be defined. In the present work we restrict ourselves to the Balanced Accuracy (BA), defined as
\begin{equation}
\text{BA} = \frac{1}{K}\sum_k\frac{ V_{kk}}{N_k}
\label{eq:BA}
\end{equation}
where $N_k=\sum_{k'} V_{k'k}$ is the number of objects in class $k$. The BA is thus simply the averaged correct prediction.

\paragraph{Training a classifier.}
Given a training set $M^\text{train}$, its associated labels $y^\text{train}$, and a possible preprocessing $X$, the aim is to choose the weight vectors $\beta_k$ in order to optimize the classifier accuracy, as evaluated for instance using the BA. However, optimizing directly the BA, which is a highly nonlinear function of the weight vectors due to taking the maximum in Eq. \eqref{Eq:argmaxy}, is generally not feasible or computationally too costly (Also for this reason a polynomial expansion, at least to high order,  is generally not feasible, and one will in fact often use PCA to reduce the number of features.)

One therefore optimizes another, simpler, metric, during training and then evaluates the performance on test data using the desired metric, e.g., the BA. The most important example is least squares regression in which one  minimizes  (averaged over the training set) the  mean squared error
\begin{equation}\label{eq:MSE}
\text{MSE} = \frac{1}{N} \sum_{i,k}
( \text{score}(x_i,k) - \delta_{y_i,k} )^2,
\end{equation}
where $\delta$ is the Kronecker delta, for which there is a well-known analytical solution for the optimal weight vectors $\beta_k$. 
In order to avoid overfitting one often puts constraints on the weight vectors $\beta_k$. For instance ridge regression and lasso regression fix their  $L_2$ and $L_1$ norms respectively:  $\vert\beta_k \vert _2 = \lambda$ or $\vert \beta_k \vert _1 = \lambda$.

In the present work we shall use data preprocessing, error metrics, and constraints on the weight vectors inspired by quantum mechanics.

\section{Quantum state discrimination}\label{Sec:3}
We start by recalling the basic concepts of quantum states and  measurements that will be used in the following.

\paragraph{Quantum states and measurements.}

A $d$-dimensional quantum state is a positive semi-definite, symmetric matrix with unit trace:
\begin{equation}
  \rho\in \mathbb{R}^{d\times d},\quad
\rho \geq 0, \quad  \quad \text{Tr}\, \rho =1\ .
\label{Eq:rho}
\end{equation}
The state $\rho$ is called a density matrix. Note that quantum mechanics is in principle over the complex numbers. But in the present context, because the input data is represented by real numbers, it will be sufficient to consider real quantum mechanics, in which all objects are vectors or matrices over the reals. 
 
A pure state is a rank $1$ density matrix. It can therefore be decomposed as $\rho = \psi \psi^T$ where $\psi$ is a $d$-dimensional real column vector normalized to $\vert \psi \vert^2=1$. In Dirac notation we would write $\rho =\vert \psi\rangle\langle\psi |$, where $|\psi\rangle$ is the ket associated to state $\psi$. States $\rho$ that are not pure are called mixed state.

A quantum measurement $E$, called a Positive Operator Valued Measure (POVM), with $K$ outcomes is a set of $K$ (symmetric) positive semi-definite matrices $E=\{ E_k\in\mathbb{R}^{d\times d}:  k=1,\dots , K\}$ that sum to the identity:
\begin{equation}
E_k\geq 0 \quad , \quad  \sum_k E_k=\mathbbm{1}\ .
\label{Eq:POVM}
\end{equation}
 The probability of obtaining outcome $k$ on state $\rho$ is given by the Born rule as 
\begin{equation} 
\Pr (k\vert \rho) = \text{Tr}(E_k \rho) \ .
\label{Eq:Born}
\end{equation}
The properties Eq. \eqref{Eq:POVM} ensure that these are indeed probabilities: $\Pr (k\vert \rho) \geq 0$ and $\sum_k \Pr (k\vert \rho) =1$.

\paragraph{Minimum error state discrimination.}

The problem of optimizing quantum measurements in order to learn as much  as possible about an unknown state has been extensively studied \cite{Helstrom1969,Chefles2000,Barnett2009,Bae2015} in the context of \emph{quantum state discrimination} \cite{Barnett2009}. This is a task where a unknown quantum state is drawn from a given ensemble
\begin{equation}
\mathcal{E} = \{(p_k,\rho_k): k=1,...,K\}
\label{Eq:EnsStates}
\end{equation}
where $p_k$ is the probability of selecting the state $\rho_k$. The aim is to design a $K$-outcome measurement $E=\{E_k\}$ that optimally discriminates the states of the ensemble $\mathcal{E}$, where an often used metric is the average probability $\Pr\!_\text{succ} (\mathcal{E})$ of successfully identifying state $\rho_k$ when the measurement gives outcome $k$, 
\begin{equation}
\Pr\!_\text{succ} (\mathcal{E}) = \sum_{k=1}^Kp_k
\Pr (k \vert \rho_k) =\sum_{k=1}^Kp_k \text{Tr}(E_k \rho)\ .
\label{Eq:PrSucc_ens}
\end{equation}

Maximizing the success probability, also known as {minimum error state discrimination}, can be formulated as an instance of semi-definite programming (SDP) \cite{Vandenberghe1996}, where the maximization is taken over the POVM elements $\{E_k\}$. Since SDPs can be efficiently solved, one can (numerically) find the optimal POVM.

\paragraph{Pretty good measurement.}
There exists an approximate solution to the minimum error state discrimination problem known as the Pretty Good Measurement (PGM):
\begin{equation}\label{Eq:PGMdef}
E_k = \rho^{-1/2} p_k \rho_k \rho^{-1/2}\,,
\end{equation}
where 
\begin{equation}
\rho = \sum_k p_k \rho_k
\end{equation}
is the average density matrix of all the states. 

One easily checks that Eq. \eqref{Eq:PGMdef} defines a valid POVM, i.e. that Eqs. \eqref{Eq:POVM} are satisfied. The probability of obtaining outcome $k'$ on input $\rho_k$ is then
\begin{equation}
\Pr (k'\vert \rho_k) = p_{k'} \text{Tr}(\rho^{-1/2}  \rho_{k'} \rho^{-1/2} \rho_k) \ .
\end{equation}

Note that when the matrix $\rho$ is full rank, i.e. when its null-space is \{0\}, it is invertible, and thus $\rho^{-1/2}$ is well-defined. When $\rho$ is not full rank, its inverse does not exist but one can consider instead its (Moore-Penrose) pseudo-inverse. This pseudo-inverse is most easily defined using the spectral decomposition of $\rho$, namely
$\rho = \sum_n \lambda_n e_n e_n^T$ where $\lambda_n$ are the eigenvalues and $e_n$ the corresponding eigenvectors. Then the pseudo-inverse $\rho^{-1}$ of $\rho$ is simply $\sum_{\lambda_n>0} (1/\lambda_n)\; e_n e_n^T$, where the sum runs only on the strictly positive eigenvalues. Hence the pseudo-inverse coincides with the inverse on the subspace orthogonal to the null-space of $\rho$ and acts as zero on this null-space. Similarly, 
$\rho^{-1/2}=\sum_{\lambda_n>0} (1/\sqrt\lambda_n)\; e_n e_n^T$.
In the following, we always assume that the inverse of $\rho$ is replaced by its pseudo-inverse whenever necessary. 

\paragraph{Improving state discrimination: multiple copies of states.} If several copies of the unknown state are provided, that is if the ensemble takes the form
\begin{equation}
  \mathcal{E} = \{p_k,\rho_k^{\otimes m}\}_{k=1}^K\,,
  \end{equation}
where $m$ is the number of copies provided and the superscript $\otimes m$ represents the $m$-tuple tensor product, then it is well known \cite{Peres1991} that it is easier to identify the quantum state using collective measurements acting on all copies.

\paragraph{Unambiguous state discrimination.}
Finally, a variant of state discrimination, called unambiguous state discrimination \cite{Ivanovic1987,Dieks1988,Peres1988}, involves the possibility to obtain an inconclusive \emph{I don´t  know} outcome. Conditioned on getting a conclusive outcome, it then becomes possible in some cases to perfectly discriminate between the states $\rho_k$ (this is the case for instance, when one must discriminate between two arbitrary pure states). The aim is then to come up with a strategy that will minimize the overall probability of an inconclusive outcome. 

In unambiguous state discrimination, the inconclusive outcome is represented by an additional measurement operator $E_0$. The new POVM is thus $E=\{E_k\}_{k=0}^K$. If we only consider operators corresponding to conclusive outcomes, the normalization condition is replaced by 
\begin{equation}
\sum_{k=1}^K E_k \leq \mathds{1}
\label{Eq:SubEk}
\end{equation}
and the probabilities become subnormalized
\begin{equation}
\sum_{k=1}^K
\Pr (k \vert \rho ) \leq 1\ .
\end{equation}

We note that adding an addditional $E_0$ outcome does not help for the regular task of minimum error state discrimination. Indeed, maximizing the success probability \eqref{Eq:PrSucc_ens} over subnormalized POVMs $\sum_{k=1}^K E_k \leq \mathds{1}$ yields the same value as maximizing over normalized POVMs $\sum_{k=1}^K E_k = \mathds{1}$. This is because from any 
POVM $\{E_k\}_{k=0}^K$ with an inconclusive outcome $E_0= \mathds{1} -  \sum_{k=1}^K E_k$, we can create a new POVM $\{ E'_k \}_{k=1}^K$ without it as follows:
$E'_k = E_k+ E_0/K$.   
The success probability $\Pr\!_\text{succ} $ using the measurement $\{ E'_k \}_{k=1}^K$ is larger than or equal to the success probability using the measurement 
$\{E_k\}_{k=0}^K$ [since  $\text{Tr}(E'_k \rho) =  \text{Tr}((E_k + E_0/K) \rho)
=   \text{Tr}(E_k  \rho) +    \text{Tr}( (E_0/K) \rho) \geq  \text{Tr}(E_k  \rho)$].
Hence we can always reach the maximum of the success probability within the smaller set of normalized POVMs.

\section{Quantum-inspired classification algorithms}
\label{sec:QICA}

\paragraph{Basic idea.}
From the above we see that if the data preprocessing Eq. \eqref{Eq:Preprocessing} is used to map the input $u_i$ to a classical representation of a density matrix:
\begin{equation}
u_i \to \rho_i 
\label{Eq:datatodensitymatrix}
\end{equation}
satisfying Eq. \eqref{Eq:rho}, then a quantum measurement $\{E_k\}$ (of dimension matching the dimension of $\rho_i$) can be used to assign  scores 
\begin{equation}
\text{score}(u_i,k) = \Pr ( k\vert \rho_i) = \text{Tr}( E_k \rho_i)
\label{Eq:scorerhoi}
\end{equation}
using Born's rule. Note that viewing $\rho_i$ as a vector, and using the fact that for symmetric matrices $A,B$, $\text{Tr}(AB)$ is a scalar product (the Hilbert-Schmidt product), this is
 a linear predictor function, as presented in Section~\ref{Sec:2}. The matrix elements of the POVM elements $E_k$ play the role of weight vector. 

We note the following important features. First, the conditions Eq. \eqref{Eq:POVM} obeyed by a POVM naturally restrict the elements of the weight vector. Thus the formulation contains by itself a regularization which may help avoid overfitting. Second, the approach naturally lends itself to multiclass classification, as the number of POVM elements can be freely chosen. Third, the scores correspond to probabilities (possibly subnormalized if one uses Eq. \eqref{Eq:SubEk}), which may useful in certain applications.

\paragraph{Mapping data to quantum states.}\label{SubSec:MapDQ}

In order to use this approach, the first step is to transform the data  vectors $u_i$ to quantum states. Since quantum states are normalized vectors, we need to map vectors $u_i$ in  $\mathbb{R}^{p}$ to normalized vectors. There are of course multiple ways of doing this. A first possibility is to normalize according to
\begin{equation}
u_i \to  x_i = \frac{u_i}{\sqrt{\vert u_i \vert^2}} \ ,
\label{Eq:UNorm}
\end{equation}
 where $\vert u_i \vert$ denotes the Euclidean norm ($L_2$ norm) of the vector $u_i$.
A second possibility is to use the inverse stereographic projection
\begin{equation}
u_i \to x_i = \left( \frac{\vert u_i \vert^2 -1}{\vert u_i \vert^2 +1}, \frac{2u_i}{\vert u_i \vert^2 +1} \right)
 \ .\label{Eq:UInvSP}
\end{equation}

Once the states are normalized, one creates a density matrix by the transformation
\begin{equation}
u_i \to \rho_i =x_i x_i^T \ .
\label{Eq:Urho}
\end{equation}
Note that this is similar to the transformation $u_i \to u_i\otimes u_i$ (see Eq. \eqref{Eq:PolyExp}) since $\rho_i$ contains all second degree monomials in the normalized features. 

Since quantum states are defined up to a phase (or for real quantum mechanics, up to a sign), the transformation of Eq. \eqref{Eq:UNorm} has the disadvantage that it looses the information on the sign of $u$ (once one computes the density matrices $u_iu_i^T$). It also has the disadvantage that it is ill defined if $u=0$ is the null vector. The transformation Eq. \eqref{Eq:UInvSP} keeps the information on the sign and is well defined for all $u$, at the expense of increasing the dimension by $1$. For large-dimensional inputs with many redundant features, both approaches give equivalent results. For small-dimensional inputs there may be a difference.

In addition to the transformations Eqs. \eqref{Eq:UNorm}  and \eqref{Eq:UInvSP}, one can also carry out some additional preprocessing on the input data, as those listed in Section~\ref{Sec:2}, i.e., standardization, PCA, polynomial expansion, etc.

After mapping the feature vectors to quantum states through \eqref{Eq:Urho}, one can also apply additional transformation. For instance, one can increase dimension by taking several copies of the density matrix
\begin{equation}
\rho_i \to  \rho_i^{\otimes m}\ .
\label{Eq:tensorproduct}
\end{equation}
This is analogous to the polynomial expansion Eq. \eqref{Eq:PolyExp}, and should improve performance at the expense of an exponential (in $m$) increase in the dimension of the density matrix.

One can  also reduce the dimension of the density matrix by the following procedure (closely related to PCA). Consider the average density matrix
\begin{equation}
\rho = \frac{1}{N}\sum_i \rho_i
\end{equation}
where $N$ is the number of examples in the training set,
and project onto the $d'$ largest eigenvectors of $\rho$:
\begin{equation}\label{Eq:proj}
\rho_i \to \frac{ \Pi_{d'} \rho_i \Pi_{d'}}{\text{Tr} \Pi_{d'} \rho_i}
\end{equation}
where $\Pi_{d'} $ is the projector onto the $d'$ largest eigenvectors. Note that if one combines \eqref{Eq:tensorproduct} and \eqref{Eq:proj} then one will apply \eqref{Eq:proj} after taking the tensor product \eqref{Eq:tensorproduct}. 

\paragraph{Training success metrics.} 
Let $\rho_i$  be the set of $N$ density matrices obtained after the transformations outlined above. From now on, we will denote the feature vectors (after normalisation and all other preprocessing) by $x_i$  and their dimension by $q$, so that
$\rho_i=x_i x_i^T$ (as in Eq. \eqref{Eq:Urho}).
 Let $\{E_k\}$ be a POVM. Let  the score assigned to outcome $k$ if the input is $x_i$ be given by the probabilities of the outcomes, i.e. by the Born rule Eq. \eqref{Eq:scorerhoi}. 
From these scores one can compute different success metrics that can be optimized in the training phase. We consider success metrics that are linear in the score, and depend only on which class input $x_i$ belongs to. These metrics will be parametrized by a $K \times K$ real matrix $\alpha_{k' k}$ (the weight if an input in class $k$ is assigned to class $k'$). They are given by:
\begin{eqnarray}
\text{SM}&=&
\sum_k \frac{\alpha_{k'k}}{N_k} \sum_{i\in C_k} \text{score}(x_i,k') \nonumber\\
&=&\sum_k \frac{\alpha_{k'k}}{N_k} \sum_{i\in C_k} 
 \text{Tr}( E_{k'} \rho_i) \nonumber\\
 &=& \sum_{k'k} \alpha_{k'k} 
 \text{Tr}( E_{k'} \bar \rho_k)\label{Eq:MetricQ}
 \end{eqnarray}
 where we have defined the \textit{centroids} for each class  as 
\begin{equation}
 \bar{\rho}_k=\frac{1}{N_k}\sum_{i\in C_k}\rho_i\ .
\end{equation}

Thus for any such linear function of the score, the classification problem can be identified to the state identification problem for the ensemble
\begin{equation}
\bar{\mathcal{E}} = \{p_k,\bar \rho_k\}_{k=1}^K\ ,
\label{Eq:barcalE}
\end{equation}
with weights given by $\alpha_{k'k}$. It is this identification which is at the heart of the present work.

Note that if we take the 
 $\alpha_{k'k}= \delta_{k k'}N_k/N$, then Eq. \eqref{Eq:MetricQ} 
becomes
\begin{equation}
\text{SM} = \Pr\!_\text{succ}  =\sum_k p_k\text{Tr} (E_k  \bar{\rho}_k)
\label{Eq:PSucc_1}
\end{equation}
where we recall that $p_k=N_k/N$. In this case the success metric is simply the success probability Eq. \eqref{Eq:PrSucc_ens}.

\paragraph{The SDP-C classifier.}
Given the success probability Eq. \eqref{Eq:PSucc_1} as training success metric, the first quantum-classifier we consider, which we denote SDP-C, is simply obtained by maximizing this metric over the training set, which can be cast as a SDP:
\begin{subequations}\label{SDP-PSucc}
\begin{IEEEeqnarray}{s,l'rCl}
maximize & \IEEEeqnarraymulticol{1}{l}{\Pr\!_\text{succ}(\bar{\mathcal{E}})} & & &  \label{SDP-Psucc-C1} \\
subject to & E_k \geq   0, & k &=& 1,\dots, K  \label{SDP-Psucc-C2} \\
           & \sum_{k=1}^K E_k = \mathds{1} , & & &
           \label{SDP-Psucc-C3}
\end{IEEEeqnarray}
\end{subequations}
where the maximization is taken over the POVM elements $\{E_k\}$. As pointed out above, one will not gain anything by relaxing the normalization condition to Eq. \eqref{Eq:SubEk} in place of Eq. \eqref{SDP-Psucc-C3}.

\paragraph{The SDP-$\gamma$-C classifier.}
The classifier \eqref{SDP-PSucc} tries to maximize the success probability. But it does not try to impose a large gap between the correct identification score $\Pr (k\vert \bar{\rho}_k)$ and the incorrect identification score $\Pr (k'\vert \bar{\rho}_k), k'\neq k$, which could be detrimental when computing the BA.
The following classifier, which we denote SDP-$\gamma$-C is based on the following SDP that tries to remediate this problem, by imposing such a gap:
\begin{subequations}\label{SDP-z}
\begin{IEEEeqnarray}{s,l'rCl}
maximize & \IEEEeqnarraymulticol{1}{l}{\gamma} & & & \label{SDP-z-C1} \\
subject to & E_k \geq   0, & k &=& 1,\dots, K \label{SDP-z-C2}\\
           & \sum_{k=1}^K E_k = \mathds{1} & & & \label{SDP-z-C3} \\
           & \text{Tr}\left(E_j\bar{\rho}_k\right)+\gamma\leq\text{Tr}\left(E_k\bar{\rho}_k\right), & \forall j&\neq& k, \quad \forall k \label{SDP-z-C4}
\end{IEEEeqnarray}
\end{subequations}
where the maximization is taken over the POVM elements $\{E_k\}$. 

Note that  using subnormalization Eq. \eqref{Eq:SubEk} in place of Eq. \eqref{SDP-z-C3} could lead to an improvement.

\paragraph{Linear Programming based classifier (LP-C).}
\label{Sub:LP-C}

The classifier \eqref{SDP-z} aims to maximize the probability to successfully identify the correct  class by imposing a gap between the (average) probabilities of correctly versus incorrectly identifying the class. This approach can be generalized to other linear predictors, which do not necessarily have a quantum interpretation. 

As illustration we consider here an approach which can be formulated as a Linear Program (LP). Let $x_i\in \mathbb{R}^q$ be some function of the input data as in Eq. \eqref{Eq:Preprocessing}, and the score be a linear predictor as in Eq.  \eqref{Eq:scorebeta_k}. Let $\bar x_k$ be the centroid vector for class $k$:
\begin{equation}
\bar x_k = \frac{1}{N_k}\sum_{i\in C_k} x_i\ .
\end{equation}

We consider the following optimization problem:
\begin{subequations}\label{LP-z}
\begin{IEEEeqnarray}{s,l'rCl}
maximize & \IEEEeqnarraymulticol{1}{l}{\gamma} & & & \label{LP-z-C1} \\
subject to & \vert \beta_k \vert_\infty \leq \Lambda, & k &=& 1,\dots, K \label{LP-z-C2}\\
           & \beta_j^T \bar{x}_k+\gamma\leq \beta_k^T \bar{x}_k, &\forall  j &\neq& k,\quad \forall k\label{LP-z-C3}
\end{IEEEeqnarray}
\end{subequations}
where the maximization is taken over the weights $\beta_k$, $\vert \beta_k \vert_\infty$ denotes the sup-norm, i.e. the maximum element of $\beta_k$, and $\Lambda$ is a meta-parameter that must  be optimized. 

Note that Eq. \eqref{LP-z-C2} corresponds to $qK$ linear constraints on the coefficients of the vectors $\beta_k=(\beta_k^{(1)},...,\beta_k^{(q)})$ expressing the fact that all coefficients must be bounded by
\begin{equation}
-\Lambda \leq \beta_k^{(j)}\leq \Lambda\quad \forall k\ ,\ \forall j\ .
\end{equation}
Thus the optimization problem \eqref{LP-z} is a Linear Program in $qK$ variables, with 
$2 qK + K(K-1)$ constraints. We denote this classifier as LP-C.

If we take in place of the $x_i$'s the vectors formed by the elements of the density matrices $\rho_i$, and $\Lambda=1$, then the optimization problem \eqref{LP-z} is a relaxation of the SDP \eqref{SDP-z}. This motivates its study in the present context, as it will be faster to solve this LP than the SDPs of the previous classifiers.

\paragraph{Pretty Good Measurement Classifier (PGM-C).} We can use as classifier the Pretty Good Measurement based on the ensemble $\bar{\mathcal{E}}$, where each POVM element $E_k$ is given by
\begin{equation}
E_k=p_k\rho^{-1/2}\bar{\rho}_k\rho^{-1/2}
\label{Eq:PGM}
\end{equation}
where $\rho=\sum_k p_k\bar{\rho}_k$. We denote this classifier as PGM-C. We study it in more detail in the next section.

Note that when $\rho$ is not invertible (which happens when the vectors $x_i$ do not span the full space $\mathbb{R}^q)$ and has to be replaced by its pseudo-inverse, the elements $E_k$ no longer sum to the identity. Since the pseudo-inverse is projecting on the subspace orthogonal to the null-space of $\rho$, they only sum to the identity restricted to this subspace and to zero elsewhere. Accordingly, the scores
\begin{equation}
  \text{score}(z,k) = \text{Pr}(k|\rho(z)) = \text{Tr}(E_k\,\rho(z)) = z^T E_k z  = z^T \Pi_S\, E_k\, \Pi_S z 
\end{equation}
computed on a test vector $z$ are subnormalized and equal to those obtained by computing the probabilities on the projection $\Pi_S z$ of $z$ on the subspace $S$ spanned by the vectors $x_i$. 

\paragraph{Mean Square Error Classifier (MSE-C).}
We experimented with another quantum-inspired classifier, the MSE-C, which is based on the minimization of the mean square error
\begin{subequations}
  \begin{IEEEeqnarray}{s,l'rCl}
  maximize & \IEEEeqnarraymulticol{1}{l}{\frac{1}{N}\sum_{i,k} \Bigg(\mathrm{Tr}(E_k\rho_i)-\delta_{y_i,k}\Bigg)^2} & & &  \label{SDP-Psucc-C1_B} \\
  subject to & E_k \geq   0, & k &=& 1,\dots, K  \label{SDP-Psucc-C2_B} \\
             & \sum_{k=1}^K E_k = \mathds{1} , & & &
             \label{SDP-Psucc-C3_B}
  \end{IEEEeqnarray}
  \end{subequations}
  We do not discuss it further here, because it is very slow compared to other algorithms, although it achieves performances similar to the PGM-C.

\paragraph{Tensor Product classifiers.} For all the above classifiers, we can define the classifier on the original density matrices $\rho_i$, or on their tensor product, following the transformation Eq. \eqref{Eq:tensorproduct}. 

As illustration, for the PGM on $m$ copies, we have 
\begin{eqnarray}
\bar{\rho}_k^{(m)} &=& \frac{1}{N_k}\sum_{i\in C_k}\rho_i^{\otimes m}\nonumber\\
\rho^{(m)}&=&\sum_kp_k\bar{\rho}_k^{(m)}
\end{eqnarray}
and then substitute into Eq. \eqref{Eq:PGM} to obtain the corresponding measurement.

We denote tensor product classifiers with the superscript $\ \otimes{m}$, for instance the PGM classifier on $m$ tensor products is noted
PGM$^{\otimes m}$-C.

\section{The Pretty Good Measurement and the Kernel Trick}
\label{Sec:PGMandKT}
\paragraph{PGM and the Gram matrix.}
We show how one can implement the Pretty Good Measurement using an approach similar to the Kernel Trick used in Support Vector Machines. This is particularly useful in order to compute the PGM on multiple tensor products of the states $\rho_i$. Indeed using the Kernel Trick, the computational hardness will scale as a power of the number of training examples $N$, rather than as a power of the Hilbert space dimension. This allows one to efficiently study the PGM as the number of copies $m$ increases and even, as we shall show, in the limit $m\to \infty$.

As before, we write the states as $\rho_i= x_i x_i^T$ and their average as
\begin{equation}
\rho = \frac{1}{N}\sum_{i=1}^N x_i x_i^T
\end{equation}
where $x_i\in \mathbb{R}^{q} $ are normalized column vectors, and $q$ is the dimension of the Hilbert space. Note that $q$ may be smaller or (much) larger than $N$, the number of training examples.

We now introduce the $N\times N$ Gram matrix
\begin{equation}
G_{ij} = x_i^T x_j\,,
\end{equation}
which is symmetric and semi-definite positive. It is full rank if the $x_i$ are linearly independent (a hypothesis we do not make).
We define the $N\times N$ symmetric matrix $P$ by
\begin{equation}\label{eq:pij}
P_{ij} = x_i^T \rho^{-1/2} x_j\,,
\end{equation}
where as usual $\rho^{-1/2}$ is the pseudo-inverse square root of $\rho$. We can then write \cite{hausladen1996classical, montanaro2019pretty}
\begin{equation}
P^2 = N G
\label{Eq:P2NG}
\end{equation}
which follows from
\begin{eqnarray}
(P^2)_{ik} &=& \sum_j x_i^T \rho^{-1/2} x_j x_j^T \rho^{-1/2} x_k\nonumber\\
&=& x_i^T \rho^{-1/2} (N \rho) \rho^{-1/2} x_k =N x_i^T x_k\ .
\end{eqnarray}
Note that we can rewrite Eq. \eqref{Eq:P2NG} as
\begin{equation}
P_{ij} = \sqrt{N} (G^{1/2})_{ij}
\label{Eq:P2NG-2}
\end{equation}
which is the form we shall use below.

Next we notice that we have the following expression for $\Pi_S$, the projector on the subspace $S$ spanned by the $x_i$'s:
\begin{equation}\label{eq:lemma1}
  \Pi_S = \sum_{i,j} x_i \left(G^{-1}\right)_{ij}\, x_j^T \ .
\end{equation}
Indeed, considering the $q \times N$ matrix $M$ having the $x_i$'s as columns (the data matrix defined at the beginning of
Section \ref{Sec:2} for the $u_i$'s), we have to find the projection on the column space of $M$ of an arbitrary vector, say $b$.
This amounts to minimizing the squared Euclidean  distance $\vert Ma - b\vert^2$, i.e. to solving for $a$ the normal equations 
$M^TMa= M^Tb$. Hence the projection $Ma$ of $b$ on $S$ is given by $Ma=M(M^TM)^{-1} M^Tb$. 
Noticing that $M^TM$ is the Gram matrix $G$,
we get Eq. (\ref{eq:lemma1}).

Consider now a new test vector $z\in \mathbb{R}^{q} $ on which we want to implement the PGM.
The POVM elements of the PGM are given by Eq. \eqref{Eq:PGM}, which we re-express as
\begin{equation}
E_k = \frac{1}{N}\sum_{i\in C_k} \rho^{-1/2} x_i x_i^T \rho^{-1/2}\ .
\end{equation}
Therefore the probability of outcome $k$ is
\begin{eqnarray}
\Pr (k\vert z) &=& z^T E_k z\nonumber\\
&=& \frac{1}{N}\sum_{i\in C_k} z^T \rho^{-1/2} x_i x_i^T \rho^{-1/2} z\nonumber\\
&=&\frac{1}{N}\sum_{i\in C_k} z^T \Pi_S \rho^{-1/2} x_i x_i^T \rho^{-1/2} \Pi_S z\ .
\end{eqnarray}
Inserting \eqref{eq:lemma1} we find
\begin{eqnarray}
  \Pr (k\vert z) &=& \frac{1}{N}\sum_{i\in C_k}  \sum_{j,k,l,m}
z^T x_j\, (G^{-1})_{jk}\, x_k^T \rho^{-1/2} x_i\, x_i^T \rho^{-1/2} x_l\,  (G^{-1})_{lm }\, x_m^T z\nonumber\\
&=& \frac{1}{N}\sum_{i\in C_k}  \sum_{j,k,l,m}
w_j^T (G^{-1})_{jk} P_{ki} P_{il}  (G^{-1})_{lm} w_m \nonumber\\
&=& 
\sum_{i\in C_k}  \sum_{j,m}  w_j^T (G^{-1/2})_{ji} (G^{-1/2})_{im} w_m\,,
\label{Pky1}
\end{eqnarray}
where we defined
\begin{equation}
  w_j = x_j^T z
\end{equation}
and used \eqref{eq:pij} and \eqref{Eq:P2NG-2}. If we denote by $\Pi_k$ the $N\times N$ matrix that projects onto the indices $k$ belonging to $C_k$:
\begin{eqnarray}
(\Pi_k)_{ij} &=& 1\mbox{ if } i=j \in C_k\nonumber\\
&=& 0 \mbox{ otherwise}\,,
\end{eqnarray}
Eq. \eqref{Pky1} can be rewritten as
\begin{equation}
\Pr (k\vert z) = w^T G^{-1/2} \Pi_k G^{-1/2} w\,.
\label{Pky2}
\end{equation}
Equation \eqref{Pky2} expresses the probability of finding outcome $k$ entirely in terms of vectors of size $N$ and matrices of size $N\times N$ which can be computed from the scalar products of $x_i$ and $x_j$, and of $x_i$ and $z$. This re-expression is thus similar to the Kernel Trick and can be used in the same way. We denote the classifier obtained by using the PGM and the Kernel Trick by kPGM-C.

\paragraph{Multiple tensor products.}\label{subsec:multproducts}
We consider the PGM and the Kernel Trick in the case where the states are  $m$-fold tensor products of the original states.
Thus the states are
$\rho_i^{(m)}=\rho_i^{\otimes m}$, and $x_i^{(m)} = x_i^{\otimes m}$. As a consequence the scalar products become
\begin{eqnarray}
\left( G^{(m)} \right)_{ij} =x_i^{(m)T} x_j^{(m)} &=&  \left( x_i^{T} x_j \right)^m
= \left( G \right)_{ij}^m\ ,
\nonumber\\
\left( w^{(m)} \right)_{i} =x_i^{(m)T} z^{(m)} &=&  \left( x_i^{T} z \right)^m
= \left( w \right)_{i}^m\ ,
\end{eqnarray}
that is one takes the $m$th power element-wise.

Denote
\begin{equation}
\epsilon = \max_{i\neq j} \vert x_i^{T} x_j \vert \ .
\end{equation}
If we assume that the $x_i$ are all different, which will be the case if there is even very slight noise or measurement error when constructing the normalized vectors $x_i$, then $\epsilon$ is strictly less than $1$: $0\leq\epsilon <1$. We can then write
\begin{equation}
G_{ij}=\delta_{ij} + \epsilon_{ij}
\end{equation}
where $\epsilon_{ii}=0$ and $\vert \epsilon_{ij}\vert \leq \epsilon <1$. Therefore
\begin{equation}
G^{(m)}_{ij} = \delta_{ij} +  \epsilon_{ij}^m,
\end{equation}
that is the Gram matrix tends exponentially fast (in $m$) towards the identity matrix. 
Consequently,  when the number $m$ of copies tends towards infinity, we have 
\begin{eqnarray}
\lim_{m\to \infty} \Pr (k\vert y) &=& \lim_{m\to \infty}  w^T  \Pi_k w
\nonumber\\
&=&\lim_{m\to \infty} \sum_{i\in C_k}\left (x_i^T z\right)^{2m}\nonumber\\
&\simeq & \lim_{m\to \infty} \max_{i\in C_k} \left (x_i^T z\right)^{2m}\,
\label{Eq:PGMlargem}
\end{eqnarray}
where in the last line in Eq. \eqref{Eq:PGMlargem} we have supposed that there is a unique $i$ that maximizes the right hand side.

Notice that $ \max_{i\in C_k} \left (x_i^T z\right)^{2m} =  
\max_{i\in C_k} \vert x_i^T z\vert$. Thus in the limit of a large number $m$ of copies, the PGM becomes extremely simple. If we wish to use it as a classifier in this limit, we proceed as follows: compute the scalar products $x_i^T z$ for all $i$, find the largest in absolute value, and assign $z$ to the corresponding class. 
This is similar to what is done for $k$-Nearest Neighbour (kNN) classifiers (in our case $k=1$) when using the inner product as the similarity measure, i.e. the correlation coefficient of the new example with the elements of the training set. Obviously this classifier is somewhat simplistic.

Equation \eqref{Eq:PGMlargem} also shows that in the limit $m\to \infty$ the probabilities $\Pr(k\vert z)$ tend exponentially fast to zero. This means that in numerical implementations there will be a limit on the number of copies $m$ that can be studied, the limit being reached once the probabilities $\Pr(k\vert z)$ become comparable to the numerical precision. We will see this occur in the examples below.

It is also worth noting that even if the use of multiple copies leads to better discrimination, this does not necessarily imply better classification performances when the PGM is used. We observe this clearly in the limit $m \to \infty$, where in the higher-dimensional feature space, different feature vectors tend to become decorrelated and mutually orthogonal. The new example $z$ also becomes orthogonal to all $x_i$'s in the training set and can no longer be classified since all probabilities of Eq. (\ref{Eq:PGMlargem}) tend to zero, except if $z$ coincides with one of the $x_i$'s, in which case the probability that $z$ belongs to the same class as $x_i$ remains equal to one in the limit $m \to \infty$. Hence in such limit, we get a perfect classification accuracy on the training set, i.e. `in-sample', which disappears for an `out-of-sample' test set.

\section{Numerical experiments}\label{Sec:Results}

In this section we apply the classifiers we have introduced to two datasets: MNIST, in Section \ref{subsec:MNIST}, and MNIST-1D in Section \ref{subsec:MNIST1D}. 

The classifiers SDP-C, SDP-$\gamma$-C with or without inconclusive outcome (that is, with the condition $\sum_k E_k = \mathbbm{1}$ or $\sum_k E_k \leq \mathbbm{1}$) give very similar results. Hence, for simplicity we only report the results for SDP-$\gamma$-C without any inconclusive measurement. We also report results for
LP-C and PGM-C. 
For these classifiers we use either one copy of the quantum state, or two copies which we denote with the superscript $^{\otimes 2}$.
For the PGM-C we implement the Kernel Trick to study arbitrary numbers $m>2$ of copies.

In order to benchmark our results, we also use two well-known classifiers on one or two copies:
ridge regression (denoted by RR-C)  which is least squares with $L_2$ regularization \cite{Kennedy2008,Gruber2017}  \footnote{As implementations of this classifier, we use the Python Sci-kit learn \cite{Pedregosa2011} function \textit{RidgeClassifier}. We  used the ``auto'' option (the algorithm chooses the solver based on the type of data). We also used fit\_intercept and copy\_X (to avoid overwriting the data). The Python ridge implementation solves the multiclass case in a one-vs-all approach. Finally, the ridge regularization parameter was chosen by cross-validation using the RidgeCV function from Python.} 
and multinomial logistic regression (denoted by LR-C), which is extensively documented e.g. in \cite{Bishop2006,Hilbe2009}.  \footnote{As implementations of this classifier, we use the Python Sci-kit learn  function \textit{LogisticRegression}. The multinomial logistic implementation  is a genuinely multiclass algorithm. We used the solver ``lbfgs'' (default). By default, the algorithm uses $L_2$ regularization.}

We also compared our result to a simple quantum-like classifier, which we denote the \emph{Quantum Nearest Centroid Classifier (QNC-C).} The QNC-C classifier takes the density matrix 
$\rho_z= zz^T$ 
of a new feature vector $z$, and assigns the score for that matrix belonging to label $k$ via the trace distance to the corresponding centroid,
\begin{equation}
 \text{score}(z,k) = \frac{1}{2} \text{Tr} \Big(\sqrt{(\bar{\rho}_k-\rho_z)^T(\bar{\rho}_k-\rho_z)}\Big) \ .
\end{equation}

\subsection{MNIST dataset}\label{subsec:MNIST}

\subsubsection{Comparison of different classifiers}

We first test our results on the 10-class well-known MNIST dataset of images of handwritten digits \cite{Lecun1998}. The data vectors are high-dimensional, each vector consisting in the pixel intensity (grey-scale) values for an image containing $28\times 28 = 784$ pixels.

The classification procedure goes as follows. We use an iterative process where, at each iteration, we begin by randomly shuffling the rows of the dataset of 70000 training samples, and we split it into one half that serves as training set and one half that serves as test set (although we have observed that using a training set of size $7.000$ is sufficient to reach the values reported here with the PGM-C). We use PCA to reduce the number of features to $5$, $10$, $20$ or $50$. After PCA, we apply  Eq. \eqref{Eq:UNorm} to normalize the vectors. We then apply one of the classifiers and repeat the whole procedure. The reported errors are the standard deviations of these trials. We iterate ten times.

In Table  \ref{tab:tab1a} we present the BAs (Eq. \ref{eq:BA}) obtained using the different approaches. Note that when the computations become impractical, we have added a N/A (not available) to the corresponding table cell.
A first glance shows that the quantum-inspired classifiers (LP-C, SDP-$\gamma$-C, PGM-C) compare rather favourably to the standard classifiers used as benchmarks (RR-C and LR-C). 

Looking at the results in more detail, we see that the best classifiers are the logistic regression (LR$^{\otimes 2}$-C), the ridge regression (RR$^{\otimes 2}$-C) and the PGM (PGM$^{\otimes 2}$-C), using two copies. It seems important, as least for this dataset, to classify based on the tensor product of the feature vector with itself or equivalently on the quantum state (density matrix), rather than on linear combinations of the features (as in RR-C). 

If we compare the quantum inspired classifiers (LP-C, SDP-$\gamma$-C, PGM-C, QNC-C) amongst themselves, we see that while for small number of features the best classifier varies, for 20 or more features it is the PGM-C that does best.

\begin{table}[H]
\begin{center}
\scalebox{0.55}{
 \begin{tabular}{||c|c|c|c|c|c|c|c|c|c|c|c|c||} 
 \hline
 features & RR-C & RR$^{\otimes 2}$-C & LR-C & LR$^{\otimes 2}$-C & LP-C & LP$^{\otimes 2}$-C & SDP-$\gamma$-C & SDP$^{\otimes 2}$-$\gamma$-C & PGM-C & PGM$^{\otimes 2}$-C & QNC-C \\ [0.5ex] 
\hline
 5 & $56.49\ \pm\ 0.37$ & $65.72\ \pm\ 0.25$ & $65.56\ \pm\ 0.16$ & \BlackCell{$71.94\ \pm\ 0.16$} & $53.56\ \pm\ 0.28$ & $59.18\ \pm\ 0.43$ & $41.50\ \pm\ 0.20$ & $51.94\ \pm\ 0.28$ & $49.61\ \pm\ 0.31$ & $54.81\ \pm\ 0.19$ & $54.37\ \pm\ 0.04$ \\ 
 \hline
 10 & $70.38\ \pm\ 0.24$ & $84.59\ \pm\ 0.34$ & $78.91\ \pm\ 0.14$ & \BlackCell{$90.52\ \pm\ 0.23$} & $69.90\ \pm\ 0.28$ & $75.11\ \pm\ 0.22$ & $57.95\ \pm\ 0.57$ & $64.33\ \pm\ 0.41$ & $70.59\ \pm\ 0.29$ & $79.49\ \pm\ 0.17$ & $73.33\ \pm\ 0.02$ \\ 
 \hline
 20 & $80.26\ \pm\ 0.20$ & $93.69\ \pm\ 0.16$ & $86.80\ \pm\ 0.12$ & \BlackCell{$95.83\ \pm\ 0.17$} & $76.34\ \pm\ 0.15$ & $77.48\ \pm\ 0.22$ & $74.52\ \pm\ 0.29$ & N/A & $84.61\ \pm\ 0.23$ & $92.10\ \pm\ 0.13$ & $83.13\ \pm\ 0.02$ \\ 
 \hline
 50 & $84.29\ \pm\ 0.14$ &\BlackCell{$96.87\ \pm\ 0.14$}& $90.17\ \pm\ 0.14$ & \BlackCell{$96.85\ \pm\ 0.16$} & $78.56\ \pm\ 0.21$ & $80.34\ \pm\ 0.25$ & $85.56\ \pm\ 0.15$ & N/A & $90.48\ \pm\ 0.23$ & N/A & $85.94\ \pm\ 0.01$ \\ 
 \hline
\end{tabular}
}
\end{center}
\caption{Balanced Accuracies for the MNIST dataset in percentage, for different classifiers. We use PCA to reduce the number of features (left column). N/A indicates that we were not able to carry out the computation. Errors are the standard deviation due to trials taken by randomizing and splitting the data vectors ten times. The best results for each number of features are shaded with grey.}
\label{tab:tab1a}
\end{table}

To better understand the difference between the two quantum measurement based classifiers, SDP-$\gamma$-C and PGM-C, we compare in Table \ref{tab:tab7a} the performance of the two classifiers using 3 different metrics: the success probability $\Pr\!_\text{succ}$ (Eq. \ref{Eq:PrSucc_ens}), the BA (Eq. \ref{eq:BA}), and the MSE (Eq. \ref{eq:MSE}). We see that  SDP-$\gamma$-C , which optimizes a quantity linear in the probabilities, has a significantly higher success probability $\Pr\!_\text{succ}$, which is a metric also linear in the probabilities. On the other hand, the PGM-C performs better on nonlinear metrics such as the BA or the MSE.

\begin{table}[H]
\begin{center}
\scalebox{0.7}{
 \begin{tabular}{||c|c|c|c|c|c|c|c|c|c|c|c|c||} 
 \hline
 classifier & $\Pr\!_\text{succ}^\text{train}$ & $\text{BA}^\text{train}$ & MSE$^\text{train}$ & $\Pr\!_\text{succ}^\text{test}$ & BA$^\text{test}$ & MSE$^\text{test}$ \\ [0.5ex] 
 \hline\hline
 SDP-$\gamma$-C & $40.05\pm\ 0.15$ & $86.24\pm\ 0.37$ & $2.76\ \pm\ 0.02$ & $39.54\pm\ 0.11$ & $85.49\pm\ 0.40$ & $2.81\pm\ 0.05$ \\ 
 \hline
 PGM-C & $31.47\pm\ 0.12$ & $91.38\pm\ 0.11$ & $1.74\ \pm\ 0.00$ & $31.09\pm\ 0.03$ & $90.34\pm\ 0.33$ & $1.78\pm\ 0.05$\\
 \hline
\end{tabular}
}
\end{center}
\caption{Comparison of training/test success probability $\Pr\!_\text{succ}$, BA and MSE for the MNIST dataset, keeping 50 features after PCA, for SDP-$\gamma$-C and PGM-C.}
\label{tab:tab7a}
\end{table}

In order to better understand the differences in performance reported in Table  \ref{tab:tab1a}  
we present in Table \ref{tab:tab1c}  the number of trainable parameters per class for each classifier for a data vector $x=(x_1,\dots,x_q)^T$ with $q$ features.

For a linear classifier as in Eq.~(\ref{Eq:scorebeta_k}), the number of trainable parameters (for each class $k$) is given by the size of the feature vector. Classifiers which directly use $x$ as input have thus $t_q=q$ trainable parameters. This is the case of LP-C, RR-C and LR-C. Note that constraints on the linear classifier could reduce by $O(1)$ the number of trainable parameters, as for instance the normalization condition (\ref{SDP-Psucc-C3}); we do not take this into account here.

For LP$^{\otimes 2}$-C, RR$^{\otimes 2}$-C and LR$^{\otimes 2}$-C, we used as data vector $(x,x\otimes x)$, which has $q+q(q+1)/2$ distinct elements. Hence these classifiers have $t_q=q+q(q+1)/2$ trainable parameters per class. 

For the SDP-C and PGM-C, the number of trainable parameters per class correspond to the non-redundant elements of the density matrix, i.e., $t=q(q+1)/2$. Note that in the case of the PGM-C, there is no actual training to do and $t$ measures the number of parameters of the POVM elements $E_k$ (which determines the complexity of computing these POVM elements). Finally, for SDP$^{\otimes 2}$-C and PGM$^{\otimes 2}$-C, we have $t=\frac{q'(q'+1)}{2}$ with $q'=\frac{q(q+1)}{2}$.

One would expect 
the performance 
to increase with the number of trainable parameters. This intuition is not always respected, and we see that while the number of trainable parameters increases dramatically for some of the classifiers, this is not reflected in the increase in performance  for classification of new out-of-sample examples.

\begin{table}[H]
\begin{center}
\scalebox{0.7}{
 \begin{tabular}{||c|c|c|c|c|c|c|c|c|c|c|c|c|c|c|c||} 
 \hline
 features & RR-C & RR$^{\otimes 2}$-C & LR-C & LR$^{\otimes 2}$-C & LP-C & LP$^{\otimes 2}$-C & SDP-$\gamma$-C & SDP$^{\otimes 2}$-$\gamma$-C & PGM-C & PGM$^{\otimes 2}$-C & QNC-C \\ [0.5ex] 
 \hline\hline
 5 & 5 & 20 & 5 & 20 & 5 & 20 & 15 & 120 & 15 & 120 &  15  \\ 
 \hline
 10 & 10 & 65 & 10 & 65 & 10 & 65 & 55 & 1540 & 55 & 1540 & 55   \\ 
 \hline
 20 & 20 & 230 & 20 & 230 & 20 & 230 & 210 & N/A & 210 & 22155 & 210 \\ 
 \hline
 50 & 50 & 1325 & 50 & 1325 & 50 & 1325 & 1275 & N/A & 1275 & N/A & 1275 \\
 \hline
\end{tabular}
}
\end{center}
\caption{Number of trainable parameters for each classifier.}
\label{tab:tab1c}
\end{table} 

\subsubsection{kPGM-C}

\begin{figure}[H]
\centering
\includegraphics[width=80mm,scale=1]{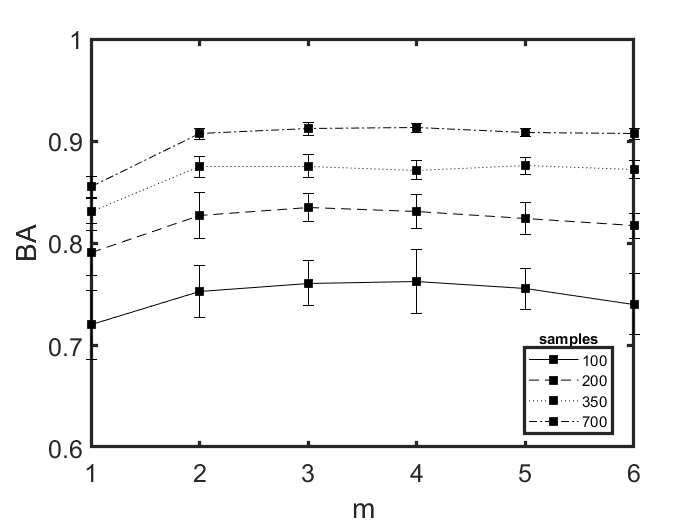}
\caption{Balanced Accuracies (BA) for different numbers of copies $m$, using kPGM-C, for MNIST dataset training set sizes of 100, 200, 350 and 700 images. The kPGM-C was tested on the rest of the dataset. The number of features is not reduced using PCA, and is therefore equal to 784.}
  \label{fig3a}
\end{figure}

Implementing the PGM using the Kernel Trick (denoted kPGM-C) allows one to study the performance of the PGM classifier on the raw unprocessed images as one increases the number of copies, without increasing the computational time.
However to implement the kPGM classifier, the size of the training set must  be reduced. In Fig. \ref{fig3a}, we give the BAs for different numbers of training examples: 100, 200, 350 and 700 images. The kPGM-C was tested on the rest of the dataset. When inverting the Gram matrix $G$, we set very small eigenvalues to zero via a threshold parameter; this amounts to using a so-called ``regularized'' pseudo-inverse where the threshold serves as regularization parameter \footnote{To do so, we use the ``pinv'' function from Matlab, and its parameter ``tol''. We perform leave-one-out cross validation on this parameter.}. We see that the BA increases with the size of the training set. We also see that increasing the number of copies $m$ from 1 to 2 results in a significant increase in BA, while subsequent increases in $m$ do not change significantly the BA. The optimal value of $m$ ranges from $3$ to $5$ depending on the number of copies, with no clear pattern emerging (after taking into account  error bars there is no clear optimal value of $m$).

\begin{figure}[H]
\centering
\includegraphics[width=80mm,scale=1]{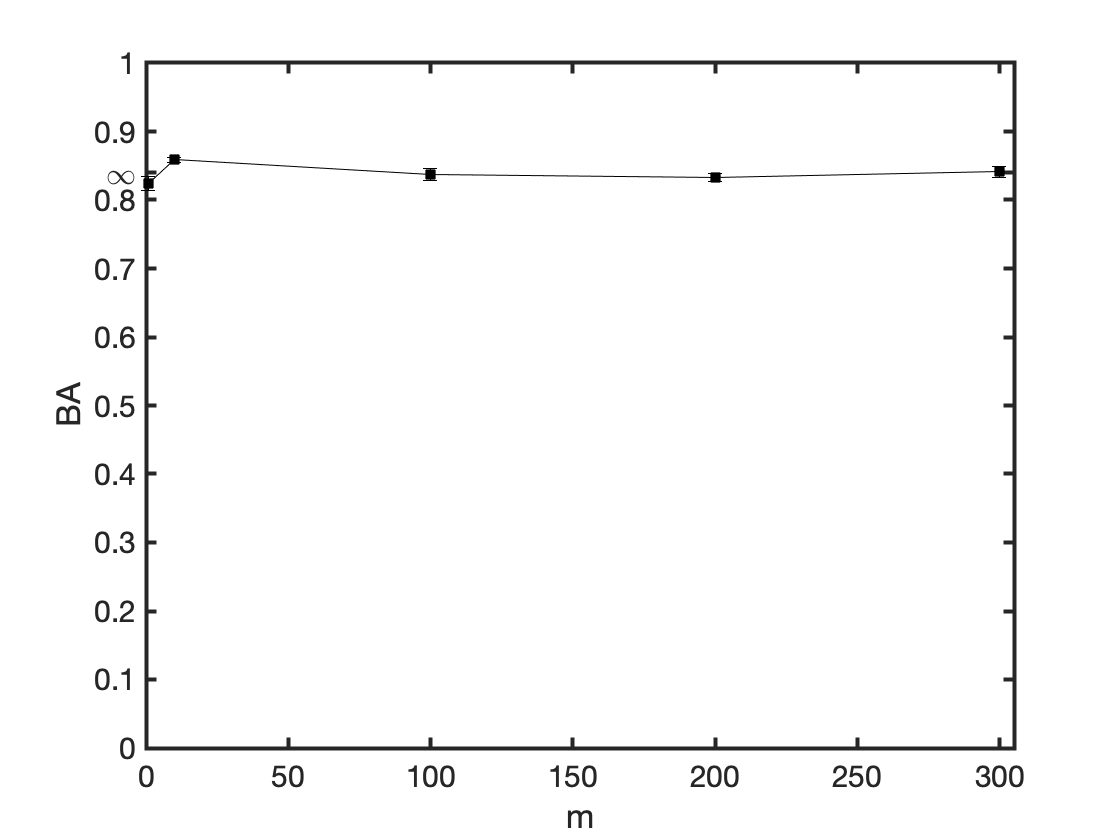}
\caption{Balanced Accuracies for different numbers of copies $m$, using kPGM-C, for a MNIST dataset size of 350 images. The plateau reached in Fig. \ref{fig3a} prolongs itself for much larger values of $m$ (up to the point where the probabilities become smaller than the numerical precision). The sign $\infty$ on the vertical axis represents the asymptotic value for $m \to \infty$.}
\label{fig:fig3c}
\end{figure}

In Fig. \ref{fig:fig3c}, we show the BA for a wider range of $m$ values. We see that the plateau reached for small values of $m$ in Fig. \ref{fig3a} prolongs itself for much larger values of $m$. When $m$ exceeds approximately $300$, the outcome probabilities become so small that Matlab effectively treats them as zeros and we have consequently cut the horizontal axis at this value.
The behavior of this curve is of course strongly dependent on the number of images used for training (350 in the figure). 
The BA of the classifier in the limit $m \to \infty$ as governed by Eq. (\ref{Eq:PGMlargem}) is given by $83.95 \pm 0.72 \%$
and is indicated by an $\infty$ sign on the vertical axis.

\subsection{MNIST-1D dataset}\label{subsec:MNIST1D}

\subsubsection{Comparison of different classifiers}

We also test our classifiers on a more recently introduced dataset, the MNIST-1D \cite{Greydanus2020}. It consists in a smaller dataset, with a total of 5000 samples with 40 features. The dataset was introduced for several reasons, including that it shows more diverse performance for linear, nonlinear and neural network classifiers, while retaining the interesting properties of the original MNIST dataset.

We train on 80$\%$ of the dataset, and test on the remaining 20$\%$ of the sample, i.e. on 1000 images. 
We apply dimensionality reduction using PCA. Results are reported in Table \ref{tab:tab4a}.

Using a tensor product of the original data improves the results. Both the RR and LC classifiers perform best, reaching a BA of close to $50\%$. 
In the case of the PGM-C, we can limit ourselves to about 25 features and still obtain a nearly optimal BA of approximately $30\%$. On the other hand LP-C and QNC-C reach BA of approximately $20\%$, while the SDP-$\gamma$-C reaches a BA significantly below $20\%$. The difference of performance of the different classifiers is very clear using this dataset.

\begin{table}[H]
\begin{center}
\scalebox{0.55}{
 \begin{tabular}{||c|c|c|c|c|c|c|c|c|c|c|c|c||} 
 \hline
 features & RR-C & RR$^{\otimes 2}$-C & LR-C & LR$^{\otimes 2}$-C & LP-C & LP$^{\otimes 2}$-C & SDP-$\gamma$-C & SDP$^{\otimes 2}$-$\gamma$-C & PGM-C & PGM$^{\otimes 2}$-C & QNC-C \\ [0.5ex] 
\hline
 5 & $21.18\ \pm\ 1.27$ & $26.93\ \pm\ 1.28$ & $23.38\ \pm\ 1.19$ & \BlackCell{$28.18\ \pm\ 1.02$} & $12.18\ \pm\ 1.01$ & $20.00\ \pm\ 1.12$  & $13.24\ \pm\ 0.96$ & $13.49\ \pm\ 0.87$ & $16.18\ \pm\ 0.78$ & $18.11\ \pm\ 1.02$ & $15.00\ \pm\ 0.01$ \\ 
 \hline
 10 & $22.26\ \pm\ 1.32$ & $33.50\ \pm\ 0.82$ & $25.06\ \pm\ 1.09$ & \BlackCell{$36.66\ \pm\ 1.05$} & $14.97\ \pm\ 1.29$ & $20.88\ \pm\ 1.06$ & $14.50\ \pm\ 0.95$ & N/A & $20.87\ \pm\ 1.00$ & $25.21\ \pm\ 0.97$ & $19.31\ \pm\ 0.02$ \\ 
 \hline
 20 & $23.70\ \pm\ 0.42$ & $46.85\ \pm\ 0.75$ & $27.71\ \pm\ 0.91$ & \BlackCell{$49.53\ \pm\ 1.06$} & $15.99\ \pm\ 0.98$ & $21.23\ \pm\ 0.70$ & $14.94\ \pm\ 1.10$ & N/A & $27.17\ \pm\ 1.09$ & $33.82\ \pm\ 0.70$ & $20.10\ \pm\ 0.01$ \\ 
 \hline
 40 & $24.77\ \pm\ 0.47$ & $47.29\ \pm\ 1.59$ & $27.49\ \pm\ 1.29$ & \BlackCell{$49.11\ \pm\ 1.24$} & $16.70\ \pm\ 1.05$ & $19.74\ \pm\ 0.84$ & $16.24\ \pm\ 1.44$ & N/A & $29.57\ \pm\ 0.92$ & N/A & $18.58\ \pm\ 0.02$ \\ 
 \hline
\end{tabular}
}
\end{center}
\caption{Balanced Accuracies for multiclass classification of the MNIST-1D dataset, in percentage. We use PCA to reduce the number of features. The best result for each number of features is shaded with grey.}
\label{tab:tab4a}
\end{table}

In Table \ref{tab:tab7b} we compare the performance of SDP-$\gamma$-C and PGM-C on one copy ($m=1$) using 3 different metrics: the success probability $\Pr\!_\text{succ}$, the BA, and the MSE. As for the case of the MNIST dataset, we see that  SDP-$\gamma$-C , which optimizes a quantity linear in the probabilities, has a significantly higher success probability $\Pr\!_\text{succ}$, 
a metric linear in the probabilities. On the other hand the PGM-C performs better on nonlinear metrics such as the BA or the MSE.

\begin{table}[H]
\begin{center}
\scalebox{0.7}{
 \begin{tabular}{||c|c|c|c|c|c|c|c|c|c|c|c|c|c||} 
 \hline
 classifier & $\Pr\!_\text{succ}^\text{train}$ & $\text{BA}^\text{train}$ & MSE$^\text{train}$ & $\Pr\!_\text{succ}^\text{test}$ & BA$^\text{test}$ & MSE$^\text{test}$ \\ [0.5ex] 
 \hline\hline
 SDP-$\gamma$-C & $14.80\pm\ 0.10$ & $25.86\pm\ 0.71$ & $13.59\ \pm\ 0.20$ & $11.99\pm\ 0.16$ & $16.36\pm\ 0.84$ & $15.47\pm\ 0.86$ \\ 
 \hline
 PGM-C & $11.97\pm\ 0.02$ & $45.50\pm\ 1.48$ & $10.57\ \pm\ 0.39$ & $11.41\pm\ 0.03$ & $29.04\pm\ 1.11$ & $13.57\pm\ 0.45$ \\
 \hline
\end{tabular}
}
\end{center}
\caption{Comparison of training/test success probability $\Pr\!_\text{succ}$, BA and MSE for the MNIST-1D dataset, 40 features, for SDP-$\gamma$-C and PGM.}
\label{tab:tab7b}
\end{table}

\subsubsection{kPGM-C}

The Pretty Good Measurement Classifier now with the Kernel Trick is applied to the MNIST-1D dataset, for different numbers of training samples. No dimensionality reduction is used, i.e. the measurement is applied to the full image comprising 40 features. Similarly to the previous dataset, we perform cross validation on the threshold parameter for computing the (regularized) pseudo-inverse of the matrix $G$.

\begin{figure}[H]
\centering
\includegraphics[width=80mm,scale=1]{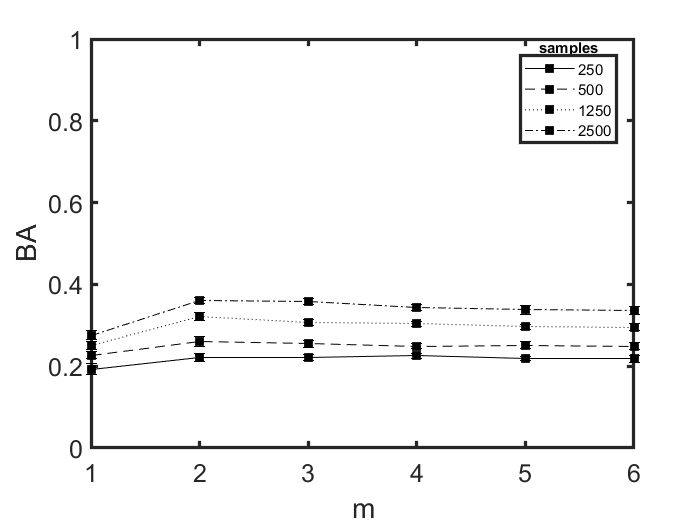}
\caption{Balanced Accuracies for different numbers of copies $m$, using kPGM-C, for MNIST-1D training set sizes of 250, 500, 1250, and 2500 images. The kPGM-C was tested on the rest of the dataset. The number of features is not reduced using PCA, and is therefore equal to 40.
\label{fig6a}}
\end{figure}

In Fig. \ref{fig6a}, we give the BAs for different training set sizes: 250, 500, 1250 and 2500 images. The kPGM-C was then tested on the rest of the dataset. We see that the BA increases with the number of training samples. We also see that the optimal number of copies (on the full dataset) is  $m=2$ except for the smallest data set when it is $m=4$ ( due to the error bars on the BA, these optimal values for $m$ are approximate).

\begin{figure}[H]
\centering
\includegraphics[width=80mm,scale=1]{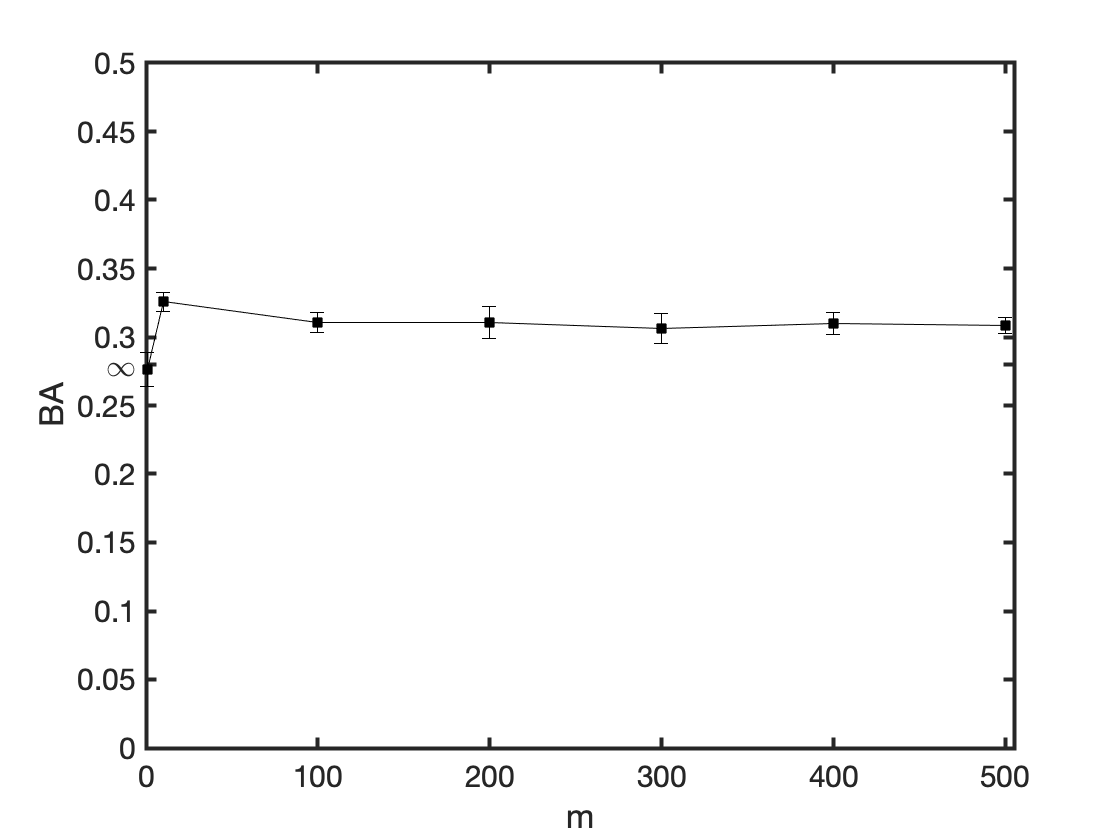}
\caption{Balanced Accuracies for different numbers of copies $m$, using kPGM-C, for a MNIST-1D training set of 1250 images. For $m$ larger than approximately $500$ copies, the outcome probabilities become so small that Matlab effectively treats them as zero. The sign $\infty$ on the vertical axis represents the asymptotic value for $m \to \infty$.}
\label{fig:fig6c}
\end{figure}

In Fig. \ref{fig:fig6c}, we show the BA for a wider range of $m$ values, with a training set of 1250 images. We see that the plateau reached for small values of $m$ in Fig. \ref{fig6a} prolongs itself for much larger values of $m$. When  $m$ exceeds approximately $500$ the outcome probabilities become so small that Matlab effectively treats them as zeros, and we have consequently cut the horizontal axis at this value. The asymptotic value for the BA in the limit $m \to \infty$ is given by $27.98 \pm 0.69 \%$
and is indicated by a sign $\infty$ on the vertical axis.

\section{Conclusion}\label{Sec:Conclusion}

 In the present article, we have proposed machine learning algorithms for classification inspired by an analogous problem from quantum information processing, namely the quantum state discrimination problem. Such a setting has been considered before in the literature  \cite{Gambs2008,Tiwari2018, Sergioli2018,Sergioli2019, giuntini2021quantum, Giuntini2023},  albeit not as extensively as in the present paper. 

These quantum-inspired classifiers have several advantages. First, they are naturally multiclass because quantum measurements can have multiple outcomes. On the other hand, most traditional classification algorithms are binary (at least in their most natural formulation). This is obviously useful for many applications. 

Second, quantum-inspired classifiers output a probability of belonging to each class. We have not used this feature explicitly in the present work. But it would be highly relevant in a Bayesian framework in which one has a priori probabilities that the inputs belong to specific classes. One would then use the output of the quantum-inspired classifiers to compute an a posteriori (i.e. after classification) probability to belong to each class.

 Third, by construction, these classifiers incorporate a natural safeguard against overfitting because 
quantum measurements are described by unit-trace positive operators, 
 and  therefore all elements of the classifier are bounded in absolute value. 
This should be contrasted with, e.g.  least squares regression for which one uses ridge or lasso regularization, or  
the Linear Programming based classifier Eq. \eqref{LP-z} which contains a meta-parameter ($\Lambda$) which plays the role of regulator and needs to be carefully tuned.  

We have found that in most cases the best quantum-inspired classifier is the Pretty Good Measurement. The PGM is particularly interesting because it has all the advantages enumerated above, and is directly constructed from the data, without the need for any numerical optimization. For this reason it is extremely fast to construct. We have also shown that the PGM could be implemented using the Kernel Trick, which makes the computational cost 
scale with the number of examples, rather than with the dimension of the Hilbert space. This allows us to study its performance when the number of copies of the quantum state is large (up to several hundred), which to our knowledge has never been done before. We find that the performance is largely flat as we vary the number of copies, with an optimum for a small number (2 or 3) of copies.

In general taking tensor products of artificial copies of the object quantum states  is expected  to be beneficial for classification task, as pointed out in 
 \cite{giuntini2021quantum}.
To see this, suppose that for each number $m$ of copies, we optimize over a set $\Sigma^{m}$  of POVMs; and suppose that these sets have the property that if $\{E_k^{(m)}\}$ is a POMV on $m$ copies that belongs to 
$\Sigma^{m}$, then the POVM  $\{{E'_k}^{(m+1)}=E_k^{(m)} \otimes \mathbbm{1} \}$ belongs to $\Sigma^{m+1}$. The POVM $\{{E'_k}^{(m+1)}\}$ acting on $m+1$ copies of the object quantum state will have exactly the same outcome probabilities as the POVM $\{E_k^{(m)}\}$ acting on $m$ copies. Hence when this hypothesis is satisfied, increasing the number of copies cannot decrease the efficiency of the classifier.

Interestingly, in the case of the PGM this turns out not to be the case, as we have shown in Section \ref{subsec:multproducts} and in our numerical examples. 
For a given size of the training set, there is an optimal number of copies, and increasing the number of copies then results in a decrease of the classification accuracy.
This is due to the fact that the PGM is given in closed form, and does not obey the hypothesis outlined in the previous paragraph.

In summary, we have found that quantum-inspired classifiers, and in particular the Pretty Good Measurement, have performances which compare rather favourably to standard classifiers found in numerical libraries. Given that the quantum-inspired classifiers have several advantages enumerated above, and that the standard classifiers have been studied and optimized since many years, this is very encouraging, and suggests that quantum-inspired classifiers will find  applications in machine learning.

We note that the quantum-inspired classifiers studied here could in principle be implemented on a quantum computer. These quantum implementations could possibly be interesting quantum algorithms that efficiently solve classification problems. But understanding whether this is the case is complex and goes beyond the present work. Indeed to this end one needs to understand the efficiency of the quantum algorithm: what is the size of the corresponding quantum circuit? Is it proportional to the dimension (which would be bad)? To the log of the dimension (which would be good)? To the size of the training set (probably bad)? 
Classifiers based on the PGM could be particularly cheap in terms of resources when running on a quantum computer \cite{Gilyen2022}.

Part of the answer to these questions is related to the difference between how the quantum versus the quantum-inspired classifier works. To implement the quantum-inspired classifier one computes, up to numerical precision, the probabilities $\Pr (k\vert \rho(z))$ for the different outcomes (labeled $k$) of the quantum measurement on the state $\rho(z)$. The output of the classifier depends on the relative values of these probabilities (typically we select as outcome the $k$ such that $\Pr (k\vert \rho(z))$ is largest).
On the other hand, if one were to input the quantum state 
$\rho(z)$ into a quantum circuit implementing the measurement, one would obtain as outcome a single value $k$. It may be that this quantum measurement can be implemented much more efficiently than the classical computation of the probabilities  $\Pr (k\vert \rho(z))$. However in order to determine the output of the classifier we  need to find the largest probability, i.e. $\text{argmax}_{k}\Pr (k\vert \rho(z))$. 
This can be done by repeating the whole process (state preparation followed by measurement) until one has accumulated enough statistics that one can reliably determine the largest probability. How many repetitions are required  depends on the gap between the largest $\Pr (k\vert \rho(z))$ and the second largest $\Pr (k\vert \rho(z))$. 
The number of repetitions required directly determines the efficiency of the quantum classifier.  It could be that the gap between the largest and second largest probability $\Pr (k\vert \rho(z))$ is very small, in which case many repetitions are required. In this work we do not study this gap between the largest and second largest probability. We leave the discussion of whether the quantum classifiers presented here could be implemented efficiently on a quantum computer to future work.

\section*{Acknowledgments}
EZC thanks support from the Swiss National Science Foundation through the Early PostDoc Mobility fellowship P2GEP2 188276, and the support from FCT -- Funda\c c\~ao para a Ci\^encia e a Tecnologia (Portugal)  through project UIDB/50008/2020. S.P. acknowledges support from the FRS-FNRS through the PDR T.0171.22.
S. M. acknowledges funding by the EOS project 
“Photonic Ising Machines” (project number 40007536) funded by the FWO and the FRS-FNRS. 
S.P. is a Research Director of the Fonds de la Recherche Scientifique – FNRS.

\bibliography{bib_2}

\end{document}